\documentclass[twoside,10pt]{article}
\usepackage{a4wide}
\usepackage[tbtags]{amstex}
\begin{document}

\makeatletter
\title {Poisson-Sigma Models\footnote{Talk presented by T.Schwarzweller}}
\author{Allen C. Hirshfeld\footnote{hirsh@hal1.physik.uni-dortmund.de} and 
Thomas Schwarzweller\footnote{thomas@doom.physik.uni-dortmund.de}\\
Institut f\"ur Physik - Theoretische Physik III\\
Universit\"at Dortmund\\}
\date{}

\maketitle
\numberwithin{equation}{section}
\begin{abstract}
{\small\sl We investigate the Poisson-Sigma model on the classical and quantum level. In the classical analysis we show how this 
model includes various known two-dimensional field theories. Then we perform the calculation 
of the path integral in a general gauge, and demonstrate that the derived partition function 
reduces to the familiar form in the case of 2d Yang-Mills theory. }
\end{abstract}

\section{Introduction}
The Poisson-Sigma model \cite{SS1} is a gauge theory based on a Poisson algebra, i.e. it is a 
non-linear extension of an ordinary gauge theory, which is based on a linear 
Lie algebra. The class of these models, which are based on a non-linear Lie algebra, i.e. 
a finite W-algebra or a Poisson algebra, are known in this context as 
{\em non-linear gauge theory} \cite{IK}. 
The Poisson-Sigma model associates to any Poisson structure on a finite-dimensional manifold a 
two-dimensional field theory \cite{SS2}. Choosing different Poisson structures leads to 
specific models which include most of the topological and semi-topological field theories 
which have been of interest in recent years. These include gravity models, non-abelian gauge 
theories and the Wess-Zumino-Witten model.

Because of the non-linearity of the algebra the Poisson-Sigma model involves in the 
language of gauge theories an open gauge algebra, i.e. \!the algebra closes only on-shell. 
In such cases the Faddeev-Popov method of path integral quantization fails. 
Quantization procedures which rely on the BRST symmetry of the extended action are in 
principle more powerful \cite{H}. We find that the antifield formalism of Batalin and 
Vilkovisky \cite{BV} is the most effective method to get a suitable action for the path 
integral quantization. The path integral for the Poisson-Sigma model was first discussed in 
a preliminary way by Schaller and Strobl in \cite{SS2}. In a recent paper Cattaneo and 
Felder \cite{CF} used the pertubation expansion of the path integral in the covariant gauge to 
elucidate Kontsevich's formula for the deformation quantization of the algebra of functions 
on a Poisson manifold \cite{KO}. Kummer et.al. have investigated the special case of 2d 
dilaton gravity and they have calculated the generating functional using BRST methods 
\cite{KLV}. We have investigated in a recent paper the path integral quantization of the 
Poisson-Sigma model in a general gauge and derived an almost closed expression for the 
partition function \cite{HSW}.

The article is structured as follows. In Section 2 we introduce the Poisson-Sigma model and 
show how the model reduces to known field theories. In Section 3 we construct the 
Batalin-Vilkovisky action and perform the calculation of the path integral. We also show 
how the derived partition function for the general model reduces under certain 
circumstances to the more familiar Yang-Mills case \cite{BT}. Section 4 contains some 
concluding remarks. 
\section{The Poisson-Sigma Model}
\subsection{Poisson manifolds and the model}
A {\em Poisson manifold} $(N,P)$ is a smooth manifold $N$ equipped with a Poisson structure 
$P\in \Lambda^2TN$  \cite{VA}. In local coordinates $X^i$ on $N$
\begin{equation}
P=\frac{1}{2}\;P^{ij}(X)\;\partial_i\wedge \partial_j,
\end{equation}
and $P^{ij}$ has to satisfy the condition
\begin{equation}
P^{i[j}P^{lk]}{}_{,i}=0,
\label{SN}
\end{equation}
which reflects the vanishing of the corresponding Schouten-Nijenhuis bracket for $P$ with 
itself. Here the bracketed indices denote an antisymmetric sum. In the notation of Poisson 
brackets
\begin{equation}
\{f(X),g(X)\}=P^{ij}(X)f_{,i}g_{,j}
\label{p1}
\end{equation}
and the Jacobi identity follows from Eq. (\ref{SN}):
\begin{equation}
\{f,\{g,h\}\}+{\rm cyclic}=0.
\label{p2}
\end{equation}
The Poisson bracket satisfies the Leibniz derivation rule:
\begin{equation}
\{h,fg\}=\{h,f\}g+f\{h,g\}\;,
\label{p3}
\end{equation}
where $f,g,h$ are functions on the manifold $N$. These facts are of an algebraic nature, and 
it is natural to define a {\em Poisson algebra} as an associative  commutative algebra endowed 
with a bracket that satisfies (\ref{p1}), (\ref{p2}) and (\ref{p3}). Indeed, a smooth manifold 
$N$ is called a Poisson manifold if the algebra of smooth functions is a Poisson algebra.

The splitting theorem of Weinstein \cite{WE} states that for a regular Poisson 
manifold, i.e. the Poisson tensor has constant rank, there exist so-called Casimir-Darboux 
coordinates on the Poisson manifold $(N,P)$. For $P$ degenerate there are nonvanishing 
functions $f$ on $N$ whose
Hamiltonian vector fields $X_f=f_{,i}P^{ij}\partial_j$ vanish. These functions are called 
Casimir functions. Let $\{C^I\}$ be a maximal set of independent Casimir functions. 
Then $C^I (X)={\rm const.} = C^I (X_0)$ defines a level surface through $X_0$ whose 
connected components may be identified with the symplectic leaves $S$ which constitute the 
symplectic foliation of the Poisson manifold $(N,P)$. According to Darboux's theorem there are 
local coordinates  $X^\alpha$ on $S$ such that the symplectic form $\Omega_S$ is given by
\begin{equation}
\Omega_S=dX^1\wedge dX^2 +dX^3\wedge dX^4 +\ldots.
\end{equation}
Together with the Casimir functions we then have a natural coordinate system
$\{X^I,X^\alpha\}$ on $N$ with $P^{IJ}=P^{I\alpha}=0$ and $P^{\alpha\beta}
= {\rm constant}$.

The Poisson-Sigma model is a field theory on a closed two-dimensional world sheet $M$.
First it involves a set of bosonic scalar fields $X^i$, which can be interpreted 
as mappings from the world sheet to a Poisson manifold, $X^i:M\rightarrow N$. 
In addition one needs fields $A_i$ which are one-forms on 
$M$ taking values in $T^\ast N$, i.e. one-forms on the world sheet which are simultanously the 
pullback of sections of $T^\ast N$ by the map $X(x)$, where $x$ denotes the coordinates of the 
world sheet. These fields can be seen as a non-linear extension of the gauge fields of an 
ordinary gauge theory. The action of the semi-topological Poisson-Sigma model is:
\begin{equation}
\mathcal{S}_{0}[X,A]=\int\limits_{M}\left[A_{i}\wedge{\rm d}X^{i}+\frac{1}{2}\;P^{ij}(X)
A_{i}\wedge A_{j}+\mu C(X)\right],
\label{CA}
\end{equation}
where $\mu$ is the volume form on $M$, $C(X)$ is a Casimir function and d denotes an 
exterior derivative on the world sheet $M$. Note that the Casmir function in the action 
breaks the topological nature of the theory.\\
The action is invariant under the following symmetry transformations:
\begin{equation}
\delta X^{i}=P^{ij}(X)\varepsilon_{j}\;,\;\;\;\delta A_{ i}=D_{i}^{j}\varepsilon_{j}\;,
\end{equation}
where $D_{i}^{j}= \delta_i^j{\rm d}+P^{kj}{}_{,i}A_{k}$ is the covariant derivative on $M$.\\
The equations of motion are
\begin{equation}
D_{i}^{j}A_{j}+\frac{\partial C(X)}{\partial X^{i}}=0\;\;\;{\rm and}\;\;\;
{\rm d}X^i+P^{ij}A_j=D X^i=0\;.
\end{equation}
The symmetry algebra is given by:
\begin{gather}
[\delta(\varepsilon_{1}),\delta(\varepsilon_{2})]X^{i}=P^{ji}(P^{mn}{}_{,j}\;
\varepsilon_{1n}\varepsilon_{2m})\;,\notag\\
[\delta(\varepsilon_{1}),\delta(\varepsilon_{2})]A_{i}=D_{i}^{j}(P^{mn}{}_{,j}
\varepsilon_{1n}\varepsilon_{2m})-(D X^{j})P^{mn}{}_{,ji}\;\varepsilon_{1n}\varepsilon_{2m}\,.
\end{gather}
Note that here the non-linearity of the Poisson algebra is manifested in the additonal term 
which is proportional to the variation of the action with respect to $A_i$.
\subsection{Three Examples}
Now we want to show how the Poisson-Sigma model reduces to specific 
two-dimensional field theories. This goal can be achieved by the choice of a particular 
Poisson structure on the Poisson manifold, which corresponds to a choice of the target 
manifold.\\[5mm]
{\bf Non-degenerate Poisson structure:}
the first example concerns the case of the nondegenerate Poisson structure, i.e the 
Poisson structure P has an inverse $\Omega$. This is exactly the symplectic 2-Form on the 
manifold $N$. 
It turns out that the target space is now a symplectic manifold. Note that in this case 
the only Casimir function is the trivial function $C=0$. It is possible to solve the equations 
of motion for the fields $A_i$, and one has:
\begin{equation}
A_i=\Omega_{ij}{\rm d}X^j.
\end{equation} 
Using this equation to eliminate the fields $A_i$ in the action, one gets:
\begin{equation}
\mathcal{S}_{top}=\int\limits_{M}\;\Omega_{ij}{\rm d}X^i\wedge{\rm d}X^j\;.
\end{equation}
This is exactly the action of the {\em topological Sigma model} proposed by E.Witten \cite{WI} in the 
Baulieu-Singer approach \cite{BS}.
\\[5mm]
{\bf Linear Poisson structure:}
next we consider a linear Poisson structure $P^{ij}=c^{ij}_k X^k$ on the three 
dimensional space $\mathbb{R}^3$. Because of the Jacobi identity the structure coefficients 
$c^{ij}_k$ define a {\em Lie algebra} structure on the dual space $\mathcal{G}$, 
of $N$. For this reason 
the linear Poisson structure is also called a Lie-Poisson structure on $N$.
It is not hard to see that the fields $A_i$ reduce to the ordinary gauge fields and their 
covariant derivative is now the ordinary curvature of a gauge theory:
\begin{equation}
F_i=D_i^j A_j={\rm d}A_i+\frac{1}{2}c^{kl}_i A_k\wedge A_l\;.
\end{equation}
For a linear Poisson structure there exist two different types if Casimir functions, namely 
the trivial case $C=0$ and the quadratic Casimir $C=\sum_iX^iX^i$\,.

For $C=0$ the action is given by:
\begin{equation}
\mathcal{S}_{BF}=\int\limits_{M}\;X^i\;F_i\;,
\end{equation}
and one sees that it is the action of a {\em topological BF gauge theory} \cite{BBT}.

Choosing now the quadratic Casimir yields for the action after a short calculation:
\begin{equation}
\mathcal{S}_{YM}=\int\limits_{M}\;F^i\wedge\ast F_i\;,
\end{equation}
where $\ast$ denotes the Hodge Star operator on the world sheet $M$. This is now the action of 
a {\em 2d Yang-Mills theory}.
\\[5mm]
{\bf 2-dimensional Gravity:}
as a last example we want to see what happens if we choose a non-linear Poisson structure on 
$\mathbb{R}^3$ given by:
\begin{align}
P^{ij} &=\epsilon^{ijk}u_k(X)\;,\\
{\rm with}\;\;u_a &=\eta_{ab}X^b\;,{\rm where}\;\; a,b=1,2\\
{\rm and}\,\; u_3 &=V(X^a\eta_{ab}X^b,X^3)\;.
\end{align}
Choosing the trivial Casimir function $C=0$ the action is:
\begin{equation}
\mathcal{S}=\int\limits_{M}\;\left(X^a\,\mathcal{D}_3A_a+X^3{\rm d}A_3+\frac{1}{2}V
\epsilon^{ab}A_a\wedge A_b\right)\;.
\end{equation}
If we now identify the first two components of the field $A_i$ with the {\em Zweibein} $e_a$ and 
the third with the {\em spin connection} $\omega$ we get for the action:
\begin{equation}
\mathcal{S}=\int\limits_{M}\;\left(X^a\,\mathcal{D}_{\omega}e_a+X^3{\rm d}\omega+\frac{1}{2}V
\epsilon^{ab}e_a\wedge e_b\right)
\end{equation}
Hence we get {\em two-dimensional dilaton gravity} with a potential $V$ depending on the dilaton field 
$X^3$ \cite{KS}.
\section{Quantization of the Model}
\subsection{The Batalin-Vilkovisky Action}
We use the path integral appraoch for quantization, because we are interested in the 
quantum theory for different space-time topologies. To obtain a suitable action for the path 
integral we apply the Batalin-Vilkovisky method for the Poisson-Sigma Model. 
We just point out some of the essential ingredients of this approach, for a general description 
see for example \cite{GPS}.

The first thing one has to do is to introduce ghosts $C$ for the symmetry transformations, and 
for each field $\Phi^A$ a corresponding antifield $\Phi_A^\ast$. For the Poisson-Sigma 
model one has:
\begin{equation}
\Phi^{A}=\{A_{\mu i},X^{i},C_{i}\}\;\;{\rm and}\;\;\Phi^{\star}_{A}=
\{A^{\mu i\star},X^{\star}_{i},C^{i\star}\}\;.
\end{equation}
For the fields and antifields there are two gradings, one is the form-degree with respect 
to the world sheet $M$ and the other is the {\em ghost number}:\\
\begin{center}\begin{tabular}{c|c|c|c}
${\rm gh}\setminus{\rm deg}$ & 0 & 1 & 2 \\
\hline
-2 & & & $C^{i\star}$ \\
\hline
-1 & & $A^{i\star}$ & $X^{\star}_i$ \\
\hline
0 & $X^i$ & $A_i$ & \\
\hline
1 & $C_i$ & & \\
\end{tabular}\end{center}
On the space of the fields and antifields one defines a symplectic structure using the 
{\em antibracket}. For the fields $B$ and $C$ this bracket is given by:
\begin{equation}
(B,C)=\sum_A\left[\frac{\partial B}{\partial\Phi^A}\wedge\frac{\partial C}{\partial
\Phi^{\star}_A}-(-1)^{{\rm deg}(\Phi^A)}\frac{\partial C}{\partial\Phi^{\star}_A}\wedge
\frac{\partial C}{\partial\Phi^A}\right]\;.
\end{equation}
One also defines a {\em Laplace operator}\,:
\begin{equation}
\triangle C=\sum\limits_{A}(-1)^{{\rm gh}(A)}\frac{\partial^2 C}{\partial\Phi^A\partial
\Phi^{\star}_A}\;.
\end{equation}
The required action has to fullfill the following conditions. First, for the  vanishing 
of the antifields the original classical action should be obtained. And second, the action must 
satisfy the {\em Quantum Master Equation}\,:
\begin{equation}
(\mathcal{S}_{BV},\mathcal{S}_{BV})-2\hbar i\triangle\mathcal{S}_{BV}=0\,.
\end{equation}
For the Poisson-Sigma model the extended action which satisfies both conditions is:
\begin{multline}
\mathcal{S}_{BV} = \int\limits_{M}\;\bigg[A_{i}\wedge{\rm d}X^{i}+\frac{1}{2}
P^{ij}(X)A_{i}\wedge A_{ j}+\mu C(X)+A^{i\star}\wedge D_{i}^{j}C_{j}+X^{\star}_{i}
P^{ji}(X)C_{j}\\
+\frac{1}{2}C^{i\star}P^{jk}{}_{,i}(X)C_{j}C_{k}+
\frac{1}{4}A^{i\star}\wedge A^{j\star}P^{kl}{}_{,ij}(X)C_{k}C_{l}
\bigg].
\label{ea1}
\end{multline}
Transforming now the extended action into Casimir-Darboux coordinates one gets:
\begin{multline}
\mathcal{S}_{BV} =\int\limits_{M}\Bigg[A_I\wedge{\rm d}X^I+A_{\alpha}\wedge{\rm d}X^{\alpha}+
\frac{1}{2}P^{\alpha\beta}A_{\alpha}\wedge A_{\beta}+\mu C(X^I)\\
+A^{I\star}\wedge{\rm d}C_I+A^{\alpha\star}\wedge{\rm d}C_{\alpha}+X_{\alpha}^{\star}
P^{\beta\alpha}C_{\beta}\Bigg]\;.
\label{ea2}
\end{multline}
Note that there are two essential simplifications. First, the terms which are quadratic in 
the ghosts vanish, and second the covariant derivative reduces to the normal exterior 
derivative. These two facts are essential for the nonpertubative calculation of the path 
integral.
\subsection{Gauge fixing}
In the Batalin-Vilkovisky approach the gauge fixing is incorporated by the {\em gauge fermion} 
$\Psi$ in the following manner. The unphysical antifields will be eliminated by the derivative 
of the gauge fermion with respect to the fields:
\begin{equation}
\Phi^{\star}_A=\frac{\partial \Psi}{\partial\Phi^A}\;.
\end{equation}
The ghost number of the gauge fermion has to be $(-1)$, so that additional fields are necessary.
The simplest choice is a so-called {\em trivial pair}\,: $\bar{C}_i,\;\bar{\pi}_i$ and the 
corresponding antifields. In the Casimir-Darboux coodinates the gauge fermion can be chosen 
to be:
\begin{equation}
\Psi=\int\limits_{M}\left[\bar{C}^I\chi_I(A_I)+\bar{C}^{\alpha}\chi_{\alpha}(X^{\alpha})
\right]\;.
\end{equation}
The gauge fixed action in Casimir-Darboux coordinates is given by:
\begin{multline}
\mathcal{S}_{\Psi} =\int\limits_{M}\bigg[A_I\wedge{\rm d}X^I+A_{\alpha}\wedge{\rm d}X^{\alpha}+
\frac{1}{2}P^{\alpha\beta}A_{\alpha}\wedge A_{\beta}+\mu C(X^I)\\
+\bar{C}^J\frac{\partial\chi_J(A_J)}{\partial A_I}\wedge{\rm d}C_I
+\bar{C}^{\alpha}\frac{\partial\chi_{\alpha}(X^{\alpha})}{\partial X^{\beta}}P^{\gamma\beta}
C_{\gamma}-\bar{\pi}^I\chi_I(A_I)-\bar{\pi}^{\alpha}\chi_{\alpha}(X^{\alpha})\bigg]\;.
\end{multline}
Now we have arrived at an action which can be used in the path integral, since the 
ambiguity in the path integral which occurs because of the gauge freedom, is removed by the 
incorporation of the gauge fermion.
\subsection{Calculation of the Path integral}
The path integral for the Poisson-Sigma model in Casimir-Darboux coordinates is
\begin{equation}
Z=\int_{\Sigma_\Psi} \mathcal{D}X^I\mathcal{D}X^\alpha\mathcal{D}A_{I}
\mathcal{D}A_{\alpha}\mathcal{D}C_I\mathcal{D}\bar{C}_I\mathcal{D}C_\alpha
\mathcal{D}\bar{C}_\alpha\mathcal{D}\bar{\pi}_I\mathcal{D}\bar{\pi}_\alpha
\exp \left(-\frac{1}{\hbar}\mathcal{S}_\Psi\right),
\end{equation}
where we have performed the usual Wick rotation $t=i\tau$, so that the exponent of the 
path integral is now real.
Integrating over the ghost and antighost fields yields the Faddeev-Popov determinants:
\begin{multline}
Z=\int_{\Sigma_\Psi}\!\!\mathcal{D}X^I\mathcal{D}X^\alpha\mathcal{D}A_{I}
\mathcal{D}A_{\alpha}\mathcal{D}\bar{\pi}_I\mathcal{D}\bar{\pi}_\alpha
{\rm det}\left(\frac{\partial\chi_{I}(A_{I})}{\partial A_{I}}\wedge{\rm d}
 \right)_{\Omega^{0}(M)}
{\rm det}\left( \frac{\partial\chi_{\alpha}(X^{\alpha})}{\partial X^{\gamma}}
P^{\gamma\beta}(X^{I})\right)_{\Omega^{0}(M)}\\
\times\exp\!\left(\!-\frac{1}{\hbar}\int\limits_{M}\Bigg[A_{I}\wedge{\rm d}X^{I}+A_{\alpha}
\wedge{\rm d}X^{\alpha}+\frac{1}{2}P^{\alpha\beta}A_{\alpha}\wedge A_{\beta}+\mu C(X^I)\!-\!\bar{\pi}^{I}
\chi_I(A)\!-\!\bar{\pi}^{\alpha}\chi_{\alpha}(X^{\alpha})\Bigg]\!\right),\\
{}
\end{multline}
where the subscripts $\Omega^{k}(M)$ indicate that the determinant results from an 
integration over k-forms on $M$.
The integrations over $\bar{\pi}_I$ and $\bar{\pi}_\alpha$ yield $\delta$-
functions which implement the gauge conditions. 
\begin{multline}
Z = \int_{\Sigma_\Psi} \mathcal{D}X^I\mathcal{D}X^\alpha\mathcal{D}A_{I}
\mathcal{D}A_{\alpha}\;{\rm det}\left( \frac{\partial\chi_I(A_{I})}{\partial A_{I}}
\wedge{\rm d} \right)_{\Omega^{0}(M)}{\rm det}\left(\frac{\partial\chi_\alpha(X^{\alpha})}
{\partial X^{\gamma}}P^{\gamma\beta}(X^{I})\right)_{\Omega^{0}(M)}\\
\times\exp\left(-\frac{1}{\hbar}\int\limits_{M}\Bigg[A_{I}\wedge{\rm d}X^{I}+
A_{\alpha}\wedge{\rm d}X^{\alpha}+\frac{1}{2}P^{\alpha\beta}A_{\alpha}\wedge A_{\beta}+\mu C(X^I)\Bigg]\right),
\end{multline}
where from now on the integrations extend only over the degrees of freedom 
which respect the gauge-fixing conditions. 
The integration over $A_{\mu \alpha}$ is gaussian, it yields 
\begin{multline}
Z = \int_{\Sigma_\Psi}\mathcal{D}X^I\mathcal{D}X^\alpha\mathcal{D}A_{I}\;{\rm det}\left(
\frac{\partial\chi_I(A^{I})}{\partial A_{I}}\wedge{\rm d}\right)_{\Omega^{0}(M)}
{\rm det}\left(\frac{\partial\chi_\alpha(X^{\alpha})}{\partial X^{\gamma}}
P^{\gamma\beta}(X^{I})\right)_{\Omega^{0}(M)}\\
\times {\rm det}^{-1/2}\left(P^{\alpha\beta}(X^{I})\right)_{\Omega^{1}(M)} 
\exp \left(-\frac{1}{\hbar}\int\limits_{M}\Bigg[A_{I}\wedge{\rm d}X^{I}+
\Omega_{\alpha \beta}{\rm d}X^\alpha\wedge{\rm d}X^\beta +\mu C(X^I)\Bigg]\right).\\
\end{multline}
Besides the term in the exponent the only dependence on $A_{I}$ is in the relevant 
Faddeev-Popov determinant. If we choose a gauge condition linear in $A_{I}$ this 
determinant becomes independent of the fields, and can be absorbed into a normalization 
factor. The integration over $A_{I}$ then yields a $\delta$-function for 
${\rm d}X^I$. When this  $\delta$-function is implemented the fields $X^I$ become 
independent of the coordinates $\{x\}$ on $M$. Hence the Casimir functions are 
constants. The constant modes of the Casimir coordinates $X_0^I$ count the symplectic 
leaves. The path integral is now
\begin{multline}
Z = \int_{\Sigma_\Psi} dX_0^I\mathcal{D}X^{\alpha}\; 
{\rm det}\left( \frac{\partial\chi_\alpha(X^{\alpha})}{\partial X^\gamma}
P^{\gamma\beta}(X^{I}_0)\right)_{\Omega^{0}(M)}{\rm det}^{-1/2}\left(P^{\alpha\beta}(X^{I}_0)
\right)_{\Omega^{1}(M)}\\
\times\exp \left(-\frac{1}{\hbar}\int\limits_M \Omega_{\alpha \beta}dX^\alpha\wedge dX^\beta
\right)\exp\left(-\int\limits_M \frac{1}{\hbar}\mu C(X_0^I)\right).
\end{multline}
The gauge-fixing of the fields $X^\alpha$ reduces the integral $\mathcal{D}X^\alpha$ to a sum 
over the homotopy classes of the maps:
\begin{multline}
Z = \int_{\Sigma_\Psi} dX_0^I \sum\limits_{[M\rightarrow S(X_0^I)]} 
{\rm det}\left( \frac{\partial\chi_\alpha(X)}{\partial X^{\gamma}}
P^{\gamma\beta}(X^{I}_0)\right)_{\Omega^{0}(M)}
{\rm det}^{-1/2}\left(P^{\alpha\beta}(X^{I}_0)\right)_{\Omega^{1}(M)}\\
\times\exp\left(-\frac{1}{\hbar}\int\limits_M \Omega_{\alpha\beta}dX^\alpha\wedge dX^{\beta}
\right)\exp\left(-\frac{1}{\hbar}\int\limits_M \mu C(X_0^I)\right).
\end{multline}
Since the $C(X_0^I)$ are independent of the coordinates on $M$ the
last exponent simplifies to
\begin{equation}
\exp\left(-\frac{1}{\hbar}\int\limits_M \mu C(X_0^I)\right)=\exp\left(-\frac{1}{\hbar}A_M 
C(X_0^I)\right),
\end{equation}
where $A_M$ is the surface area of $M$. The form of the path integral  
then becomes
\begin{multline}
Z = \int_{\Sigma_\Psi} dX_0^I \sum\limits_{[M\rightarrow S(X_0^I)]} 
{\rm det}\left( \frac{\partial\chi_\alpha(X)}{\partial X^{\gamma}}
P^{\gamma\beta}(X^{I}_0)\right)_{\Omega^{0}(M)}
{\rm det}^{-1/2}\left(P^{\alpha\beta}(X^{I}_0)\right)_{\Omega^{1}(M)}\\
\times\exp\left(-\frac{1}{\hbar}\int\limits_M \Omega_{\alpha\beta}dX^\alpha\wedge dX^{\beta}
\right)\exp\left(-\frac{1}{\hbar}A_M C(X_0^I)\right).
\label{pf}
\end{multline}
Note that we have now arrived at an almost closed expression for the partition function for the Poisson-Sigma model, i.e. all the functional integrations have been performed.
\subsection{The case of the linear Poisson structure}
We again consider the special case where the Poisson manifold $N=\mathbb{R}^3$, and the 
Poisson structure is linear: $P^{ij}=c^{ij}_k X^k$. Here we are interested in the case of 
the quadratic Casimir function which leads to the 2d Yang-Mills theory. The corresponding 
symplectic leaves are two dimensional spheres characterized, in the Casimir-Darboux coordinates,
 by their radius $X^I_0$. Weinstein \cite{WE} has shown that the symplectic leaves of a 
linear Poisson structure are the co-adjoint orbits of the corresponding compact, connected 
Lie group G of $\mathcal{G}$. Because the Lie algebra has three dimensions we are restricted 
here to the case were the Lie group is the group SU(2). By a theorem of Kirillov these orbits 
can in turn be identified with the irreducible unitary representations of G \cite{KI1}. 

These considerations can be used to further reduce the expression for the
path integral. Consider the homotopy classes of the maps 
$X^{\alpha}:M\longrightarrow S(X^{I}_0)$.
The Hopf theorem tells us that the mappings $ f,g:M\longrightarrow S(X^{I}_0)$ are 
homotopic if and only if the degree of the mapping $f$ is the same as the degree of $g$. 
This means that the sum over the homotopy classes of the maps $[X^{\alpha}]$ can be expressed 
as a sum over the degrees $n={\rm deg}[X^{\alpha}]$:
\begin{equation}
\sum_{[X^{\alpha}]}\longrightarrow \sum_{n\in\mathbb{Z}}\;.
\end{equation}
For  a map $f:X\longrightarrow Y$, where $X$ and $Y$ are k-dimensional oriented manifolds
and $\omega$ a k-form on $Y$, the degree of the mapping is given by 
\begin{equation}
\int\limits_{X}\,f^{\ast}\omega={\rm deg}[f]\int\limits_{Y}\omega\;.
\end{equation}
Using this formula yields:
\begin{equation}
 \int\limits_{M}\;\Omega_{\alpha\beta}{\rm d}X^{\alpha}\wedge {\rm d}X^{\beta}=n\int
\limits_{S}\Omega_{S}(X^I_0)\;,
\end{equation}
where $\Omega_{S}(X^I_0)$ is the symplectic form on the corresponding leaf $S$.
This gives for the partition function of Eq. (\ref{pf})
\begin{multline}
Z = \int_{\Sigma_\Psi}{\rm d}X^{I}_{0}\sum_{n\in\mathbb{Z}}
{\rm det}\left( \frac{\partial\chi_\alpha(X)}{\partial X^{\gamma}}
P^{\gamma\beta}(X^{I}_0)\right)_{\Omega^{0}(M)}{\rm det}^{-1/2}\left(P^{\alpha\beta}(X^{I}_0)
\right)_{\Omega^{1}(M)}\\
 \times\;\exp\left(-\,n\int\limits_{S}\Omega_{S}(X^I_0)\right)
\exp\left(-\frac{1}{\hbar}A_{M}C(X^{I}_{0})\right).
\end{multline}
The sum over $n$ yields a periodic $\delta$-function:
\begin{multline}
Z  = \int_{\Sigma_\Psi}{\rm d}X^{I}_{0}\sum_{n\in\mathbb{Z}}{\rm det}\left( 
\frac{\partial\chi_\alpha(X)}{\partial X^{\gamma}}P^{\gamma\beta}(X^{I}_0)\right)_{\Omega^{0}
(M)}{\rm det}^{-1/2}\left(P^{\alpha\beta}(X^{I}_0)\right)_{\Omega^{1}(M)}\\
\times \delta\left(\int\limits_{S}\Omega_{S}(X^I_0)-n\right)
\exp\left(-\frac{1}{\hbar}A_{M}C(X^{I}_{0})\right).
\label{per}
\end{multline}
The $\delta$-function says that the symplectic leaves must be integral. By the
identification of the leaves with the co-adjoint orbits, the orbits must also be integral. 
The fact that the orbits are integral reduces the number of the co-adjoint orbits to a 
countable set, which we label by $\mathcal{O}(\Omega)$.

We now consider the two determinants in the path integral. 
We choose the ``unitary gauge'' $\chi_\alpha(X^\alpha)=X^\alpha$,
so that $\partial \chi_\alpha(X) / \partial X^\gamma = \delta_\gamma^\alpha$,
and the two determinants have the same form.
The restriction of the scalar fields to the Casimir-Darboux coordinates $X^I$corresponds 
to the restriction of the scalar fields to the invariant Cartan subalgebra
considered by Blau and Thompson in \cite{BT2}, so we may adopt their 
argumentation concerning the powers to which the determinants occur for a manifold with
Euler characteristic $\chi(M)$. The result is a factor
\begin{equation}
{\rm det}(P^{\alpha\beta}(X^I_0))^{\chi(M)}.
\end{equation}
The determinant of a mapping equals the volume of the image of that mapping,
hence the determinant ${\rm det}(P^{\alpha\beta}(X^I_0))$ corresponds to  the 
symplectic volume of the leaf, which we denote by ${\rm Vol}(\Omega_S(X^I_0))$. 
The path integral then takes the form:
\begin{equation}
Z=\int_{\Sigma_\Psi}{\rm d}X^{I}_{0}\sum_{n\in\mathbb{Z}}{\rm Vol}(\Omega_{S}(X^{I}_{0}))^{\chi(M)}
\;\delta\left(\int\limits_{S}\Omega_{S}(X^I_0)-n\right)
\exp\left(-\frac{1}{\hbar}A_{M}C(X^{I}_{0})\right).
\end{equation}
Implementing the $\delta$-function by integrating over $X^{I}_{0}$
the sum over the mapping degrees becomes a sum over the set $\mathcal{O}(\Omega)$ of the 
integral orbits:
\begin{equation}
Z=\sum_{\mathcal{O}(\Omega)}{\rm Vol}(\Omega_{S}(X^{I}_{0}))^{\chi(M)}
\exp\left(-\frac{1}{\hbar}A_{M}C(X^{I}_{0})\right).
\end{equation}
Because of the identification of the integral orbits with the irreducible 
unitary representations this leads to a sum over the representations.
A special form of the character formula of Kirillov \cite{KI2} says that the symplectic volume 
of the co-adjoint orbit equals the dimension of the corresponding  
irreducible unitary representation. So the final form of the partition function is
\begin{equation}
Z=\sum_{\lambda} {\rm d}(\lambda)^{\chi(M)}\exp\left(-\frac{1}{\hbar}A_{M}C(\lambda)\right),
\end{equation}
where $\lambda$ denotes the irreducible unitary representation corresponding 
to the co-adjoint orbit, and ${\rm d}(\lambda)$ is the dimension of this 
representation. This is exactly the partition function for the two-dimensional 
Yang-Mills theory \cite{BT}. 
When we omit the Casimir term in the action we get just a sum over the dimensions of the 
representations, which is the correct result for the BF-theory, see e.g. \cite{BT2}.
\section{Concluding Remarks}
The Poisson-Sigma model is more than a unified framework for different topological and 
semi-topological field theories.  Due to its reformulation of the degrees of freedom of the 
theories in terms of the coordinates of a Poisson manifold it achieves a description in terms 
of the natural variables of general dynamical systems. Gauge theories, which are characterized 
by singular Lagrangians, cannot in general be described in terms of symplectic manifolds; 
the foliation which is characteristic for Poisson manifolds is neccesary.

Such a description of gauge theories allows one to discuss the quantization by a direct 
application of the techniques of deformation quantization. The connection between the 
Poisson-Sigma model and the deformation quantization was shown by Cattaneo and Felder 
\cite{CF} by calculation of the pertubation expansion in the covariant gauge.
Our nonpertubative calculation of the path integral which depends essentially on the framework 
of Poisson manifolds leads to an almost closed expression for the partition function. In the 
special case of a linear Poisson structure we were able to calculate the well known formula 
for the partition function of the SU(2) Yang-Mills theory with the help of fundamental facts of 
the representation theory of groups and algebras. We believe that further research will uncover 
ways of utilizing these structures  more thoroughly. The techniques used here should in 
principle 
be applicable in more general situations than the particular case in Section (3.4). Finally, 
an understanding of the mechanisms active in the general case could help to understand 
the structure of gauge theories in a more fundamental way.
\bigskip\\
{\bf Acknowledgement}

This work was supported in part (T.Schwarzweller) by the {\em Deutsche Forschungsgesellschaft} in 
connection with the Graduate College for Elementary Particle Physics in Dortmund.  


\begin{thebibliography}{9999}
\bibitem{SS1}{\sc P.Schaller, T.Strobl: }{\sl Poisson-$\sigma$-models: A Generalization of 2d 
Gravity-Yang-Mills Systems}, Talk delivered at the Conference on Integrable Systems, 
Dubna 1994, e-Print Archive: {\bf hep-th/9411163}
\bibitem{IK} {\sc N.Ikeda: }{\sl Two-dimensional Gravity and Nonlinear Gauge Theory},
Ann.Phys. {\bf 235} 1994, 435, e-Print Archive: {\bf hep-th/9312059}
\bibitem{SS2} {\sc P.Schaller, T.Strobl: }{\sl  Poisson Structure Induced (Topological) Field Theories in 
two Dimensions}, Mod. Phys. Lett. {\bf A9} (1994), 3129
\bibitem{H} {\sc M.Henneaux, C.Teitelboim: }{\sl  Quantization of Gauge Systems}, 
Princeton University Press, Princeton, New Jersey (1992)
\bibitem{BV} {\sc I.A.Batalin, G.A.Vilkovisky: }{\sl Gauge Algebra and Quantization},
Phys. Lett. {\bf 69B} (1977), 309
\bibitem{CF} {\sc A.S.Cattaneo, G.Felder: }{\sl A Path Integral Approach to the Kontsevich Quantization 
Formula}, e-Print Archive: {\bf math/9902090}
\bibitem{KO} {\sc M.Kontsevich: }{\sl Deformation Quantization of Poisson Manifolds I},
e-Print Archive:\linebreak {\bf q-alg/9709040}
\bibitem{KLV} {\sc W.Kummer, H.Liebl, D.V.Vassilevich: }{\sl Exact Path Integral Quantization of generic 
2d Dilaton Gravity}, Nucl.Phys {\bf B493} (1997), 491
\bibitem{HSW} {\sc A.C.Hirshfeld, T.Schwarzweller: }{\sl Path Integral Quantization of the 
Poisson-Sigma Model}, e-Print Archive: {\bf hep-th/9910178} (to appear in Ann.Phys. (Leipzig))
\bibitem{BT} {\sc A. Migdal: }{\sl Recursion Relations in Gauge Theories}, Zh. Eksper. Teoret. Fiz.
{\bf 69} (1975), 810 (Soviet Physics JETP. {\bf 42}, 413).\\
{\sc E.Witten: }{\sl On Quantum Gauge Theories in Two Dimensions}, Comm. Math. Phys. 
{\bf 141} (1991), 153\\
{\sc M.Blau, G.Thompson: }{\sl Quantum Yang-Mills Theory on Arbitrary Surfaces}, 
Int. J. Mod. Phys. {\bf A7} (1992), 3781
\bibitem{VA} {\sc I.Vaisman: }{\sl Lectures on the Geometry of Poisson Manifolds}, 
Progress in Mathematics Volume {\bf 118}, Birkh\"auser, Basel, 1994
\bibitem{WE} {\sc A.Weinstein: }{\sl The Local Structure of Poisson Manifolds}, 
J. Differential Geometry {\bf 18} (1983), 523
\bibitem{WI} {\sc E.Witten: }{\sl Topological Sigma Models}, 
Comm.Math.Phys. {\bf 118} (1988), 411
\bibitem{BS} {\sc L.Baulieu, I.Singer: }{\sl The topological Sigma Model},
Comm.Math.Phys. {\bf 125} (1989), 227 
\bibitem{BBT}{\sc D.Birmingham, M.Blau, G.Thompson: }{\sl Topological Field Theory}, 
Phys.Rept. {\bf 209} (1991), 129 
\bibitem{KS} {\sc W.Kummer, D.J.Schwarz: }{\sl General analytic Solution of $R^2$ Gravity with dynamical Torsion in 
two Dimensions}, Phys.Rev. {\bf D 45} (1992), 3628 
\bibitem{GPS} {\sc J.Gomis, J.Paris, S.Stuart: }{\sl Antibrackets, Antifields and Gauge-Theory 
Quantization}, Phys. Rept. {\bf 259} (1995), 1
\bibitem{KI1} {\sc A.Kirillov: }{\sl The Orbit Method I, Geometric Quantization}, 
in Representation Theory of Groups and Algebras, Contemporary Mathematics Volume {\bf 145}, 1993
\bibitem{BT2} {\sc M.Blau, G.Thompson: }{\sl Lectures on 2d Gauge Theories}, 
presented at the 1993 Trieste Summer School in High Energy
Physics and Cosmology, \linebreak
e-Print Archive: {\bf hep-th/9310144}
\bibitem{KI2} {\sc A.Kirillov: }{\sl Elements of the Theory of Representations}, 
Grundlehren der mathematischen Wissenschaften {\bf 220},
Springer, Berlin, 1976
\end{thebibliography}
\end{document}